\documentclass[english,aps,prx,twocolumn,showpacs,superscriptaddress,floatfix]{revtex4-1}
\usepackage[T1]{fontenc}
\usepackage[latin1]{inputenc}
\usepackage{amsmath}
\usepackage{graphicx}
\usepackage{amsmath}
\usepackage{subfig}
\usepackage{amssymb}
\usepackage{dcolumn} 
\usepackage{bm} 
\usepackage{color}
\usepackage{cleveref}

\makeatletter

\usepackage{amsfonts}
\usepackage{epsfig}
\usepackage{dsfont}

\newcommand{\be}{\begin{equation}}
\newcommand{\bea}{\begin{eqnarray}}
\newcommand{\eea}{\end{eqnarray}}
\newcommand{\ee}{\end{equation}}

\usepackage{babel}
\makeatother

\begin{document}
%

\title{A geometrical imaging of the real gap between economies of
China and the United States}

\author{Ali \surname {Hosseiny}}
\affiliation{Department of Physics, Shahid Beheshti University, G.C., Evin, Tehran 19839, Iran}
\affiliation{School of Particles and Accelerators, Institute for Research in Fundamental Sciences (IPM) \\P.O.Box 19395-5531, Tehran, Iran}

 \email{alihd22@gmail.com, Al_hosseiny@sbu.ac.ir}
\date{\today}



\begin{abstract}

 GDP of China is about 11 trillion dollars and GDP of the United States is about 18 trillion dollars. Suppose that we know for the coming years, economy of the US will experience a real growth rate equal to \%3 and economy of China will experience a real growth as of  \%6. Now, the question is how long does it take for economy of China to catch the economy of the United States. The early impression is that the desired time is the answer of the equation $11\times1.06^X=18\times1.03^X$. The correct answer however is quite different. GDP is not a simple number and the gap between two countries can not be addressed simply through their sizes. It is rather a geometrical object. Countries pass different paths in the space of production. The gaps between GDP of different countries depend on the path that each country passes through and local metric. 
To address distance between economies of China and of the US we need to know their utility preferences and the path that China passes to reach the US size. The true gap then can be found if we calculate local metric along this path. It resembles impressions about measurements in the General Theory of Relativity. 
Path dependency of aggregate indexes is widely discussed in the Index Number Theory. Our aim is to stick to the geometrical view presented in the General Relativity to provide a visiual understanding of the matter. We show that different elements in the general relativity have their own counterparts in economics. We claim that national agencies who provide aggregate data resemble falling observers into a curved space time. It is while the World Bank or international organizations are outside observers. The vision provided here, leaves readers with a clear conclusion. If China keeps its growth rate, then the economy of China should catch the economy of the United States sooner than what we expect.

\end{abstract}


\maketitle
\thispagestyle{empty}



\section{Introduction}
Suppose that you want to extrapolate the time that economy of China needs to become as big as the economy of the United States. The current size of economy of China is about 11 trillion dollars and the current size of economy of the US is 18 trillion dollars. Now, suppose that we are sure that economy of China in coming years will experience a real growth as of \%6 and economy of the US experiences a \%3 rate. Then, the question is how many years does it take for economy of China to catch economy of the US. Our impression is that we can simply find the answer through solving the equation
\bea\begin{split}\label{equation}
&11\times1.06^X=18\times1.03^X\;\;\;\;\;\; \cr&\;\;\;\;\Rightarrow X=\frac{ln(18/11)}{ln(1.06/1.03)}\approx17 \;years.
\end{split}\eea
This answer is however untrue. GDP is not a simple number. To find GDP we aggregate a set of productions through their prices. In other words we have
\bea\label{gdpintro}
GDP=\Sigma_aP_aY_a,
\eea
in which $a$ ranges for any form of final goods or services provided in a country, $Y_a$ is the quantity of each good and $P_a$ is its price. Though the output of the aggregation is a single number, this number is not a regular number. The fact is that prices are our meter to perform aggregation. Elements of this meter however vary with heterogenous rates over time.

Difficulty now arises if we aim to compare GDP of two different countries. Even for a single country we have difficulty when we aim to compare GDP of different years. Problem is as follows. Nominal GDP in the year $i$ is
\bea
GDP^i=\Sigma_aP^i_aY^i_a,
\eea
and nominal GDP in the consecutive year is 
\bea
GDP^{i+1}=\Sigma_aP^{i+1}_aY^{i+1}_a.
\eea
If we decide to compare both GDPs and find growth rate
\bea\label{nominalintro}
g_{nominal}=\frac{GDP^{i+1}}{GDP^i}-1=\frac{\Sigma_aP^{i+1}_aY^{i+1}_a}{\Sigma_aP^i_aY^i_a}-1,
\eea
we face problem. The problem is that GDPs of different years have been measured with different set of prices. As a result, inflation may result in an overestimation of the measurement. So, we should find which part of the nominal growth is real growth and which part is inflation.

The usual method is to compare the GDP of the consecutive years with the same set of prices and define the real growth as
\bea\label{realintro}
g=\frac{\Sigma_aP^{i}_aY^{i+1}_a}{\Sigma_aP^i_aY^i_a}-1.
\eea
Now, it seems that we have overcome the problem with inflation. We however have created another problem. The definition of the real growth in Eq. (\ref{realintro}) fails the circularity test. 

Let us consider a time series of prices $P^i_a$ and productions $Y^i_a$ over a period of $m$ years. For each year we find the real growth through Eq. (\ref{realintro}). If at the beginning and the end of the period production and prices are the same then it means that GDP have not grown over the period. So, we expect 
\bea
\Sigma^{i+m}_{j=i}g_j\approx 0.
\eea
This trivial requirement however is dismissed in our definition. What is the consequence? The consequence is path dependency of measurements. I have supplemented a couple of csv files. In one of them I have produced a time series of prices and production for a bi-sector economy. Two countries start with exactly initial conditions. Over a century, one of the countries experiences a real growth annually equal to $3.7\%$ and another country experiences a real growth equal to $2.2\%$. Despite this huge differences, both countries end with exactly the same conditions and exactly the same set of production and prices. Guess what! Despite their equal performance for the whole period, the governing party of the first country is proud and celebrates a century of annual growth of $3.7\%$ and the governing party of the second country is ashamed.

Dependency of measurements to the paths first attracted attention of Fisher in price index problem in 1927 \cite{Fisher}. He noticed if we aim to measure inflation then our measurement can be path dependent. Finding a proper method to measure inflation then was under attention for decades. Some influential movements in indexing were addressed by Laspeyres, Paasche,  T\"ornqvist, and Fisher. For recent works see Balk 1995 \cite{balk}, Diewert 1976 \cite{Diewert76} and 1993 \cite{Diewert93}, Feenstra and Reinsdorf 2000 \cite{feenstra2000} and Oulton 2008 \cite{oulton2008}.  

In recent years paradoxical results concerning evaluation of GDP reported by international organizations such as the World Bank have reraised attentions. In International Comparison Program (ICP) and Penn World Table, GDP of each country is evaluated through purchasing power parity (PPP) exchange rate. When we calculate GDP of different countries through PPP exchange rate it means that we have used a standard price for each good worldwide. In other words we have used the same meter to perform aggregation in Eq. (\ref{gdpintro})
 for all countries and measure the size of GDP of them. We then can find out growth of GDP of each country for a time interval. Surprisingly what we earn is different from the ones that are reported in the system of national accounts (SNA). For works concerning paradoxical results see \cite{Crawford}-\cite{prasada}. In these works both dependency of measurements on the paths of growth and as well other problems such as miss-measurements have been discussed. Paradoxical results in this direction sometimes have been called space time inconsistency (see Feenstra, Ma and Prasada Rao 2009 \cite{freenstra2009}, Feenstra, Inklaar and Timmer 2013 \cite{feenstra2013}, and Oulton 2015 \cite{oulton2014}) where by space people mean measurements in different countries.

In this paper we borrow the concept of measurement and metric in the General Relativity to understand the problem. Though path-dependency of measurements for the growth rate is an old known fact in economy and has being discussed in the Index Number Theory, it has been barely known outside of economics literature. Even in the world of economy it is not discussed in regular textbooks. The geometrical vision provided in this paper helps to clearly understand 
why GDP is not a simple number and why measurement for the growth rate is path dependent. It then can be easily visioned and understood by researches in different areas of sciences. Previously, interesting efforts have been devoted to find some invariant measures for aggregate indexes through an analysis of the gauge invariance of economic models, see Malaney and Weinstien1996 \cite{malaney}, and Smolin 2009\cite{smolin}. Our work however relies on geometrical imaging of the problem and mainly notifies some similarities between objects and concepts in the general theory of relativity and the index number theory.  

Metric in the general relativity is addressed by Einstein's action. In economics, equilibrium constraints and production functions address the relation of prices. While in the general relativity objects move along geodesics, in economics, utility preferences address the path of growth in the space of productions. We discuss that national agencies who measure GDP resemble falling observers
 in a curved space-time and have their own metric to calculate their growth rate. It is while the World Bank is an outside observer which uses a standard metric for all of these countries. Besides, we discuss that to address the gap of economy of China and of the US we
 need to know production functions and utility preferences to find their paths in the space of production. Then moving along this path we can find out the time it needs for economy of China to catch economy of the US. The overall impression is that China may catch economy of the US sooner than what we expect.

Besides problem with measurement we shortly review the cost disease phenomenon and its
 role to address a major feature of heterogeneity of dynamics of long run prices. 
In econophysics we usually have been interested in short run prices and its volatility.
 A vast range of works have been dedicated to the price in the stock market, see for 
example \cite{mantegna}-\cite{silva}. Beside stock market,
physicists along with a community of economists have devoted big efforts to the
 heterogeneous agent based models and emergent macroeconomics, see for example \cite{Fujiwaraa}-\cite{mastromatteo}. In major parts of these works short run behavior of the market is of interest. In our work however we mainly focus on the long run behavior. 

While in the general theory of relativity, metric is found through the variation of Einstein's action, in economy the long run prices are addressed through the interaction of different sectors and the hypothesis of equilibrium conditions. We shortly review production function and notify the source of heterogeneity in dynamics of prices in the context of the Baumol's cost disease, see\cite{Baumol1966}-\cite{ngai}. 

This paper is organized as follows. In the following sections we shortly review
production function and the Baumol's cost disease phenomenon. Here we notify the source of heterogeneity of dynamics of prices in the long run. 
In Sec. III we discuss path dependency of measurements in economics. Besides, we introduce
our toy model to produce a time series of prices for a bisector economy. Our time series
is supplemented so that enthusiastic reader can perform his/her own analysis on measurements
of the growth rate. In Sec. IV we provide the main core of our work. In this section
we borrow geometrical vision in the General Relativity to visualize path dependency of
measurements in macroeconomy. We as well notify correspondence of concepts in the
general relativity with the mentioned problem. In Sec. V on the basis of geometrical 
vision provided we explain how the gap between economy of China and the
United States can be addressed and conclude our paper.

\section{Dynamics of long run prices in a heterogenous world}

In neoclassical framework of economy, long run prices and its relation to the wages and interest rate can be addressed through studying production functions which we shortly review in this section.
Production function is a function that indicates output of a firm in sector $a$ as
\bea\label{productionfunction}
Q_a=Y_a(T_aL_a,K_a)
\eea
 in which $L_a$ indicates the number of employed labors, $K_a$ states the value of invested capital, $T_a$ stands for productivity, and $Q_a$ indicates the physical quantity of production in the firm. The function is an ever increasing function of its variables. Besides, it is supposed to have diminishing return to capital. In other words we expect it to be a convex up function respect to capital. Further, we require the function to have constant return to scale     
\bea\label{productionscale}
Y_a(zT_aL_a,zK_a)=zY_a(T_aL_a,K_a).
\eea

For a manager it is reasonable to hire more labor if the value of added production compensates for the extra wages or equivalently
\bea
P_a\Delta Y_a \ge W\Delta L_a
\eea
in which $P_a$ is the price of unit of production and $W$ is the wage of each labor. In equilibrium in a competitive market we expect
\bea\label{wage}
P_a\frac{\partial Y_a}{\partial L_a}=W.
\eea
Same scenario goes for capital. Renting new capital is reasonable if extra production compensates for the interest rate and depreciation of capital.
\bea
P_a\Delta Y_a \ge \Delta K_a(r+\delta_a)
\eea
in which $r$ indicates the interest rate and $\delta_a$ indicates depreciation of capital. In equilibrium we expect for all sectors
\bea\label{capital}
P_a\frac{\partial Y_a}{\partial K_a}=R_c+\delta_a,
\eea
in which $R_c$ is the net rate of return on capital. Classically it was supposed that given the price of production in each sector, Eq. (\ref{wage}) indicates the level of wage of labors in that sector. It was however notified by Baumol and Bowen in their influential paper that $W$ should be the same for all sectors and it is the price of production that should be adjusted in Eq. (\ref{wage}). The consequence of the statement of Baumol and Bowen was important. Since wage and rate of return on capital should be the same for all sectors, then equations (\ref{wage}) and (\ref{capital}) leave us with a conclusion. Sectors which have relatively higher rates of growth in productivity should loose their relative prices. It is because $Y_a$ is an ever increasing function of $L_a$ or $T_a$.  Let's compare two sectors which we call sectors $A$ and $B$. At a given time Eq. (\ref{wage}) states that 
\bea
P_A\frac{\partial Y_A(T_AL_A,K_A)}{\partial L_A}=P_B\frac{\partial Y_B(T_BL_B,K_B)}{\partial L_B}
\eea
Now, if technology or actually the value of $T_A$ substantially grows in sector $A$, then to keep the equity we need to relatively increase $P_B$. So, in a heterogeneous world, prices grow with heterogeneous rates. Baumol and Bowen thereby anticipated that prices of productions or services in the stagnant sectors of the market should grow comparing to the sectors which have high rate of growth in technology. In 1900 each labor in agriculture sector could annually harvest say 1000 kg of carrot. At the same time a teacher could teach 25 students. So, if a farmer compensated 40 kg of carrot for teaching his sun then we had a peace. In 2015 each labor may annually harvest 50000 kg of carrot. Despite growth of productivity in agriculture sector, in education sector a teacher still can teach only 25 students. So to keep balance, cost of education for a farmer's sun should be equal to 2000 kg of carrot. So, from 1900 to 2015 inflation in education sector should have been much higher than agriculture sector.

Since productivity in service sectors regularly do not grow as fast as manufacturing sectors then it means that the rate of growth of prices in service side should be much higher than industrial side. This fact is called the Baumol's cost disease phenomenon. The disease has been observed in services in recent decades. In Baumol et al. 2012 \cite{Baumol2012} many evidences have been cited. For example it mentions that according to the report of National Center for Public Policy and Higher Education in 2008 \cite{national}, since the early 80s, while prices in average had grown by 110\%, expenses for education had faced a growth as of 440\%. Besides, expenses for health care had grown as of 250\%. If we notice that service sector consists a big portion of sectors in economy, then an average of 110\% means that inflation for industrial side should have been much below this rate. Beside education and health care, higher rates of inflation has been observed in many other sectors in service side. According to the report of the U.S. Bureau of Labour Statistics 2009a the cost of funeral services has roughly doubled since 1987 to 2008\cite{usbureaub}. As another example while overall inflation has been around 3\%, lawyers fee has grown around 4.5\% between 1986 and 2008 \cite{usbureau}.

Back to our discussion over production function we notify that constant return to scale (\ref{productionscale}) means that production function can be written as
\bea
Y_a(T_aL_a, K_a)=L_aY_a(T_a,\frac{K_a}{L_a})=L_ay_a(T_a,k_a)
\eea
in which $y_a$ is production per labor and $k_a=\frac{K_a}{L_a}$ is capital per labor. Now, Eq. \ref{capital} reduces to
\bea\label{newcapital}
\frac{\partial y_a}{\partial k_a}=R_c+\delta_a, 
\eea 
and Eq.(\ref{wage}) reduces to
\bea\label{newwage}
P_ay_a=W+k_a(R_c+\delta_a).
\eea

As it can be seen $L_a$ is eliminated from equations. Though prices and capital per labor are addressed in equations (\ref{newcapital}) and (\ref{newwage}), neither $K_{a}$ nor $Y_a$ can be identified. Only given $R_c$ we can identify $K_{a}/L_a$, $Y_a/L_a$, and $P_a$. For a more precise discussion over the matter in a heterogeneous world see \cite{hosseinyintensive}.To address aggregate production  we need to know $L_a$. Distribution of labors are given from demand side. In demand side we need to know the utility preference of customers. They optimize their satisfaction or equivalently maximize a utility function such as 
\bea\label{timeutility}
U=\int^{\infty}_{t=t_0} e^{-\rho t}u(t)dt
\eea
 in which $u(t)$ is instant utility. If we forget dynamics and accumulation of capital, we may write utility as $u(Y_a-\delta_aK_a)$. This is a function that differs in different countries and cultures. People in southern Europe prefer to enjoy sitting in cafe or restaurant while Americans may enjoy more having bigger cars or bigger houses. So, even if all production functions and variables such as rate of return on capital is the same in these countries, the relative size of sectors is different. If we know the utility function then we can maximize it through variation over $L_a$. Actually we have $\Sigma_aL_a=L_t$. Lagrange method leads to equations which address $L_a$
\bea\label{lagrange}
\frac{\partial u(Y_a-\delta_aK_a)}{\partial L_a}=Constant.
\eea
Solving equations (\ref{newcapital}) to (\ref{lagrange}) theoretically we can find $L_a$ and thereby $Y_a$ and perform aggregation. This is what we will do in our toy model and measure GDP growth rates for different conditions.

Before closing the section I should notify that our discussion is rough and wont work for the short run prices at all. We ignored fluctuation of prices in the market. We as well ignored accumulation of capital which itself can be addressed in the context of endogenous growth models. We simply supposed that $R_c$ is the same for all sectors which is not true in the short run. Despite being rough in the short run, the discussed framework provides acceptable approximation for the long run relations over the prices and distribution of capital.

%
%
%
%

\section{GDP Growth Rate; a Path Dependent Measure}

It has been known since long ago that measurement of the growth rate is path dependent. Actually studies have been more concerned with price index which is almost the same problem. There is no way to define a practical measure for price index which passes a couple of tests such as circularity test. Circularity test means that when we measure a chained index over time if at the beginning and the end of a period conditions are the same then we expect our chained index be equal to one. 

Same difficulty goes for measurements for the growth rate as well. Suppose that we want to compare GDP of two different years. Nominal GDP of the $i_{th}$ year can be written as
\bea
GDP_i=\Sigma_aP_a^iY_a^i
\eea
in which $Y_a^i$ is the physical quantity of production in the sector $a$ and $P_a^i$ is the price for units of production. Aggregation should be performed over all sectors. Now, if we aim to compare this level of GDP to the one from another year say $j_{th}$ year, we can write $GDP_j=\Sigma_aP_a^jY_a^j$. We then face a problem. Prices are subject to inflation and comparing nominal GDP we may overestimate the growth of the size of GDP. If all prices grew with the same rate then we could simply divide $GDP_j$ with the size of inflation.

As it was discussed in the previous section, prices are subject to heterogeneous growth rates. As a result aggregation of the real growth in a invariant method is impossible. In 1939 Kon\"us defined a measure to be path independent \cite{konus}. He supposed that all consumers do have a utility preferences which sustains its functional form over the time and space. Even if we accept such hypothesis, it seems that estimating such function seems not to be an easy task. There has been efforts to optimize some methods that may practically evaluate Kon\"us index, see Oulton 2008 \cite{oulton2008}. In any circumstance, national agencies usually use chained  Paasche, Laspayres, Fisher or T\"ornqvist indexes nowadays which are path dependent measures and will be considered in this paper. 

To find GDP growth rate, usually GDP of the sequential years are compared with the same price base
\bea\label{growthtoy}
g=\frac{\Sigma_aP_a^iY_a^{i+1}-\Sigma_aP_a^iY_a^{i}}{\Sigma_aP_a^iY_a^{i}},
\eea
in which $Y_a^{m}$ is the production level in sector $a$ in the $m_{th}$ year and $P_a^{m}$ is the price of unit of production in this sector.

\begin{figure*}[]
\centering
  \subfloat[][]{
    \includegraphics[width=1\columnwidth]{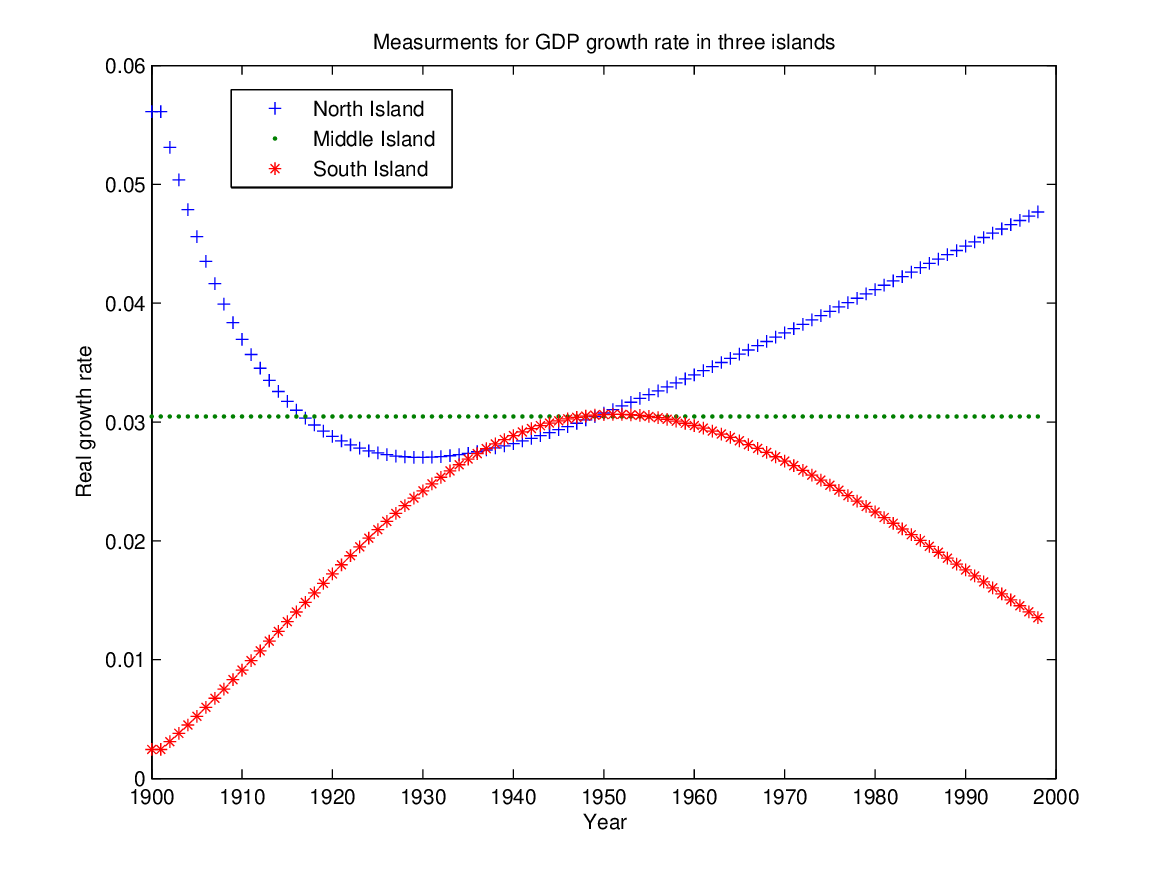}
                \label{figgeommetrygrowth}}
  \subfloat[][]{
    \includegraphics[width=1\columnwidth]{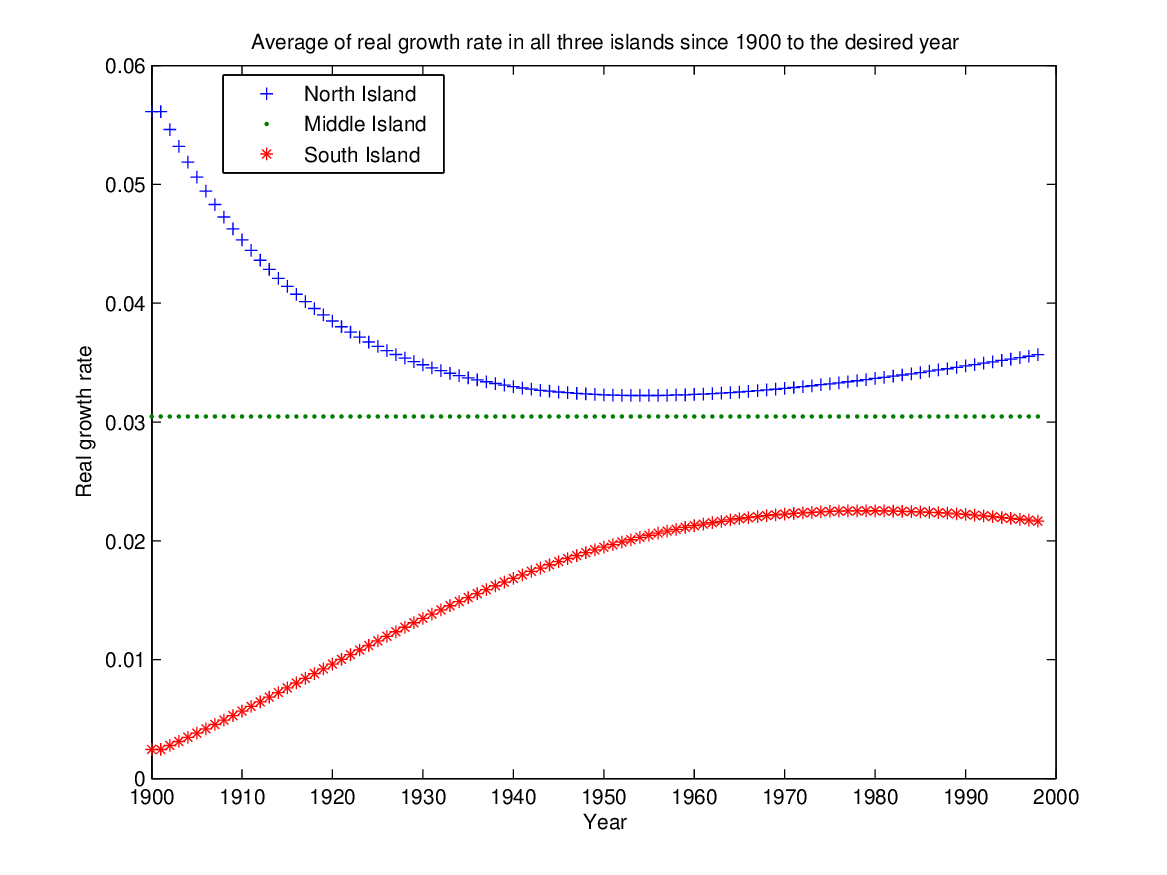}
                \label{figgeommetryavrage}}
             \caption{ a) Real growth rate in each Island. For each year the previous year has been considered as the base year. b) The average of growth of each island since 1900 to the desired year has been calculated. Each country has its own pattern of growth and thereby the average of growth should vary in middle years. The initial and final size of economies are however the same. So, we expect that the average of real growth in the end of the period to be the same for all countries. As it can be seen the average of the growth for three islands do not merge as their final conditions merge in 1998.}
\label{figgrowthes}
\end{figure*}

Now, we introduce a toy model and utilizing a chained growth rate we show that dependency of measurements when structural changes happens can be significant. We will see that for three countries with exactly initial and final conditions, perceptions for the real growth rate by national agencies range from 2\% to 3.5\%. A reader who is not willing to 
get engaged in technical discussions may escape the rest of the current section and jump to the following section where we have provided geometrical imaging of the discussion. 

Let's suppose that we have three Islands which we call South, Middle, and North Island. Each Island has a closed economy and in this closed economy they produce only one good to eat which we denote as good $A$ and provide only one service as $B$. Production functions have Cobb-Douglas form ($Y_a=T_a^{\lambda}L_a^{\lambda}N_a^{1-\lambda}$) and we set $\lambda_A=\lambda_B=2/3$. We suppose that utility function has a form as 
\bea\label{utilitymiperception}
u=(\frac{Y_A}{L_t}-N^0)(\frac{Y_B}{L_t})^{\Omega},
\eea
in which we suppose $N^0=1.6711$ and $\Omega=5$. Such utility function means that people need a minimum value of food, otherwise utility is negative. Once the minimum food which has been indicated by $N_0$ is fulfilled then people look for service $B$. When productivity is low the major portion of labors work in sector $A$. As long as economy grows then the major of labors move to the sector $B$ since $\Omega=5$. It resembles structural changes in economy. In the early decades of the twenties century we had a movement of labors from the agriculture sectors to the manufacturing sectors. In the recent decades as well we have observed substantial structural changes where economy has moved from an industrial base to a service base. Actually structural changes happens in all sectors in economy over time where each sector experiences an Engel's consumption cycle. Based on this cycle each good or service at the beginning appear as a luxury good with high income elasticity and in the end appears as necessity with a low elasticity.

\begin{figure}%
   \centering
\subfloat[][]{
\includegraphics[width=1\columnwidth]{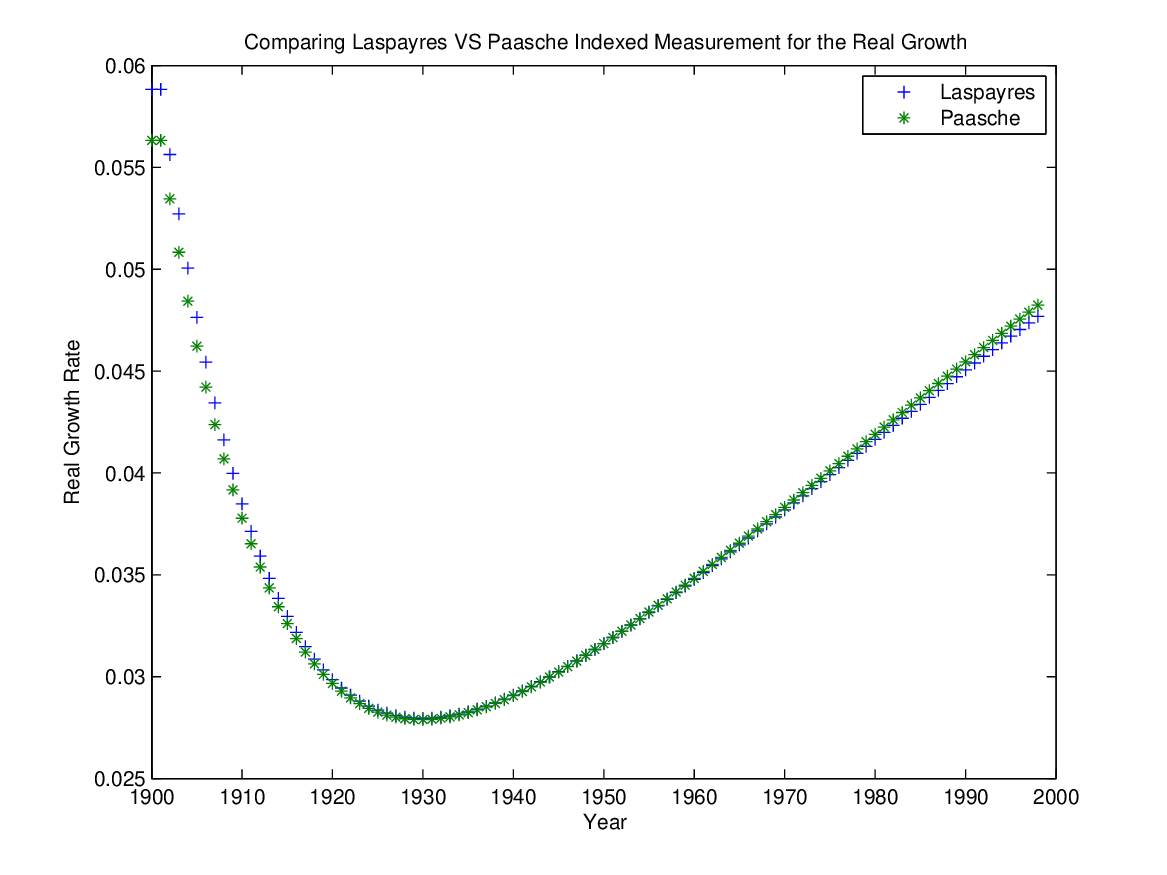}
         \label{astc}}
           \caption{ Measurement for the real growth rate based on different inflation indexes. As it can be seen a result from a Laspayres index is very close to the Paasche index in the North Island. So, choosing different indexes can not relax the problem with measurements.} 
  \label{laspayres}
 \end{figure}

In our toy model we suppose that total number of labors in each Island is 100000. Concerning productivity we suppose that in 1900 $T_A=T_B=1$ and in 1998 $T_A=T_B=18.93$. Though initial and final conditions are the same for all Islands, we suppose that in the intermediate years each island follows its own R\&D program. We suppose that in North Island in early years a big deal of attentions is devoted to the sector $A$ and productivity grows faster in this sector and in the late years productivity grows faster in sector $B$. Quantitatively we consider the following pattern for growth in productivities
\bea\begin{split}
&T^{1900+i}_A=(1+0.06*\frac{100-i}{99})*T_A^{1900+i-1}\cr&T^{1900+i}_B=(1+0.06*\frac{i+1}{99})*T_B^{1900+i-1}.
\end{split}\eea
In the Middle Island we suppose that productivity grows with a sustainable rate in both sectors
\bea\begin{split}
&T^{i}_A=(1.0305)*T_A^{i-1}\cr&T^i_B=(1.0305)*T_B^{i-1}\;\;\;\;\;\;\;\;\;\;\;\;\;\;\;\;\;\;\;\;\;\;\;\;\;\;\;\;.
\end{split}\eea
In the South Island we suppose that despite North Island in early years most efforts are devoted to the sector $B$ and productivity grows in this sector faster in early years
\bea\begin{split}
&T^{1900+i}_A=(1+0.06*\frac{i+1}{99})*T_A^{1900+i-1}\cr&T^{1900+i}_B=(1+0.06*\frac{100-i}{99})*T_B^{1900+i-1}.
\end{split}\eea
Given these rates, if we maximize utility we find distribution of labors as
\bea\begin{split}\label{labormispersepction}
&L_A=L_t\frac{\lambda_A+\Omega\lambda_BN^0/T_A(\frac{1-\lambda_A}{R_c+\delta_A})^{\frac{1-\lambda_A}{\lambda_A}}}{\lambda_A+\Omega\lambda_B},\cr&L_B=L_t-L_A.
\end{split}\eea
For depreciation we set $\delta_a=\%5.5$. We kept rate of return on capital sustainable and equal to $\%5.5$. We then ran a simple code to carry out calculation concerning GDP real growth rates for each Island. Five CSV files have been supplemented named by gdpnorth.csv, gdpmiddle.csv, and gdpsouth.csv which represent our results. Each file is a 99*4 matrix where each row represents, YA, PA, YB, and PB for of the related year from 1900 to 1998. As it can be seen in the file, initial and final level of production and prices are the same for three countries. Now, we can measure the real growth rate for these data. When we use Eq. (\ref{growthtoy}) then for the growth rate of each island we come to the results depicted in Fig. \ref{figgeommetrygrowth}. The average of the growth rate from 1900 to the desired year has been depicted in Fig. \ref{figgeommetryavrage}. As it be can bee seen as we approach 1998 all three countries approach to the same final condition and we expect the average of the real growth rate for the whole period merge to a unique value. It is while in Fig. \ref{figgeommetryavrage} we observe that in the end of the period, the average of the real growth for the whole period is \% 3.5 in North Island, \%3 in Middle Island and \%2.1 in South Island. What a surprise! While initial and final conditions are the same for all three countries, the governing parties in the North Island are proud of their performance for the past century while the governing parties in South Island should be ashamed for the actually the same results. To make sure that different indexes can not influence the paradoxical result in our toy model we considered both Laspayres and Paasche index and recalculated the real growth rate. As it can be seen in Fig \ref{laspayres} the difference is negligible. Since Fisher and T\"ornqvist indexs are between Laspayres and Paasche indexes then it means that none of these indexes can not influence the result noticeably.

Besides the mentioned csv files, two other files have been submitted as "northconstant.csv" and "southconstant.csv". Time series in these files are even more interesting than the former files. Two islands start with exactly initial conditions. One of them annually experiences a real growth around $3.7\%$ and of them experiences an annual growth around $2.2\%$ (I mean the annual growth for every year within the period and not the average of the annual growth.). Despite this annual substantial differences, the islands finally end up with exactly the same economy.

\section{Geometrical Understanding of the Growth Rate}

\begin{figure}
        \centering
  \includegraphics[width=1\columnwidth]{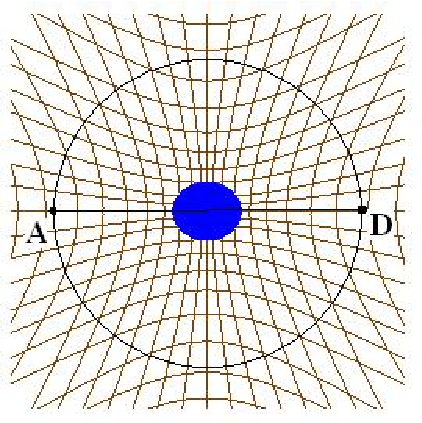}
     \caption{Near a star metric for people who are at circumference is different from who are close to the star in diameter. One astronaut moves along the circumference from A to D and measure it. Another astronaut moves along diameter from A to D to measure distance. If he is not cooked in the middle of the star and meet the other astronaut at point D surprisingly finds that for radius and circumference we have $S<2{\it{\Pi}} R$  }\label{asta}
\end{figure}

We obtained surprising results concerning measurements for the growth rate in our toy model. Now, let's recall a similar problem in physics. We suppose that we have two astronauts in two spacecrafts. They start a trip from one point in one side of a star to the other side as depicted in Fig. \ref{asta}. In Fig. \ref{asta} they start their trips simultaneously from point A. One of the spacecrafts moves along the diameter and the other spacecraft move along the circumference from A to D. They have different speeds and manage to end their trip simultaneously at point D. During their trip they measure both distance and time of their trip. When they share their measurements they observe that surprisingly the relation between radius and circumference is no longer $S=2{\it{\Pi}} R$. Beside of the distance they disagree on the duration of the trip. Neither of the astronauts has made mistake. The point is that space becomes curved by gravity. In Fig. \ref{asta} distance between each line in the grid shows the same value say one kilometer.

The story about measurements for GDP growth rate is exactly similar to the problems with astronauts. When we measure the real growth rate
\bea
g=d \log{GDP}=\Sigma_a\frac{P_a}{\Sigma_b P_bY_b} d Y_a,
\eea  
for any changes in production we use prices as our meters. Let's look at it from a geometrical point of view.

\begin{figure*}%
\centering
  \subfloat[][]{
    \includegraphics[width=.65\columnwidth]{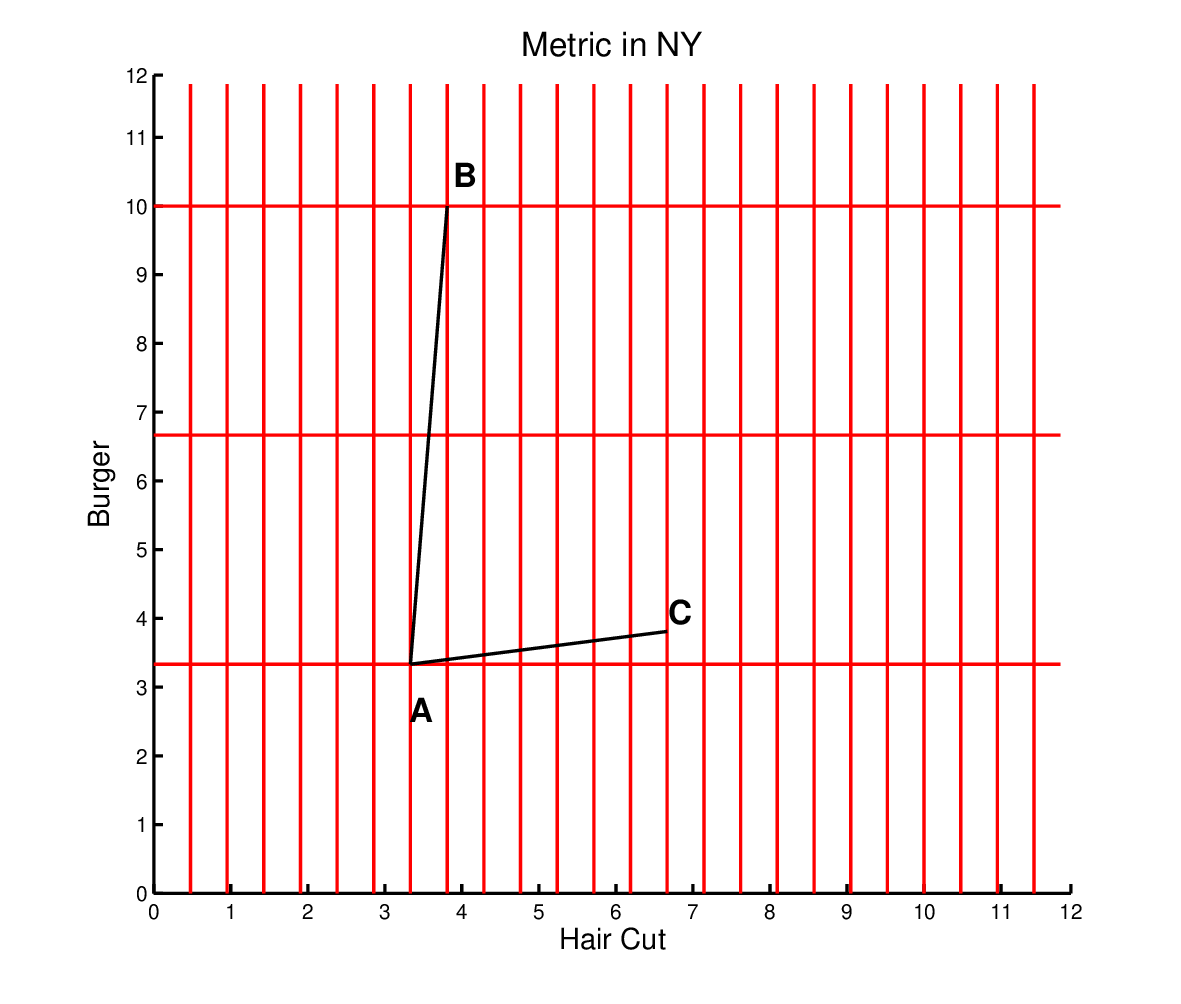}
                \label{figburgerus}}
 \subfloat[][]{
    \includegraphics[width=.65\columnwidth]{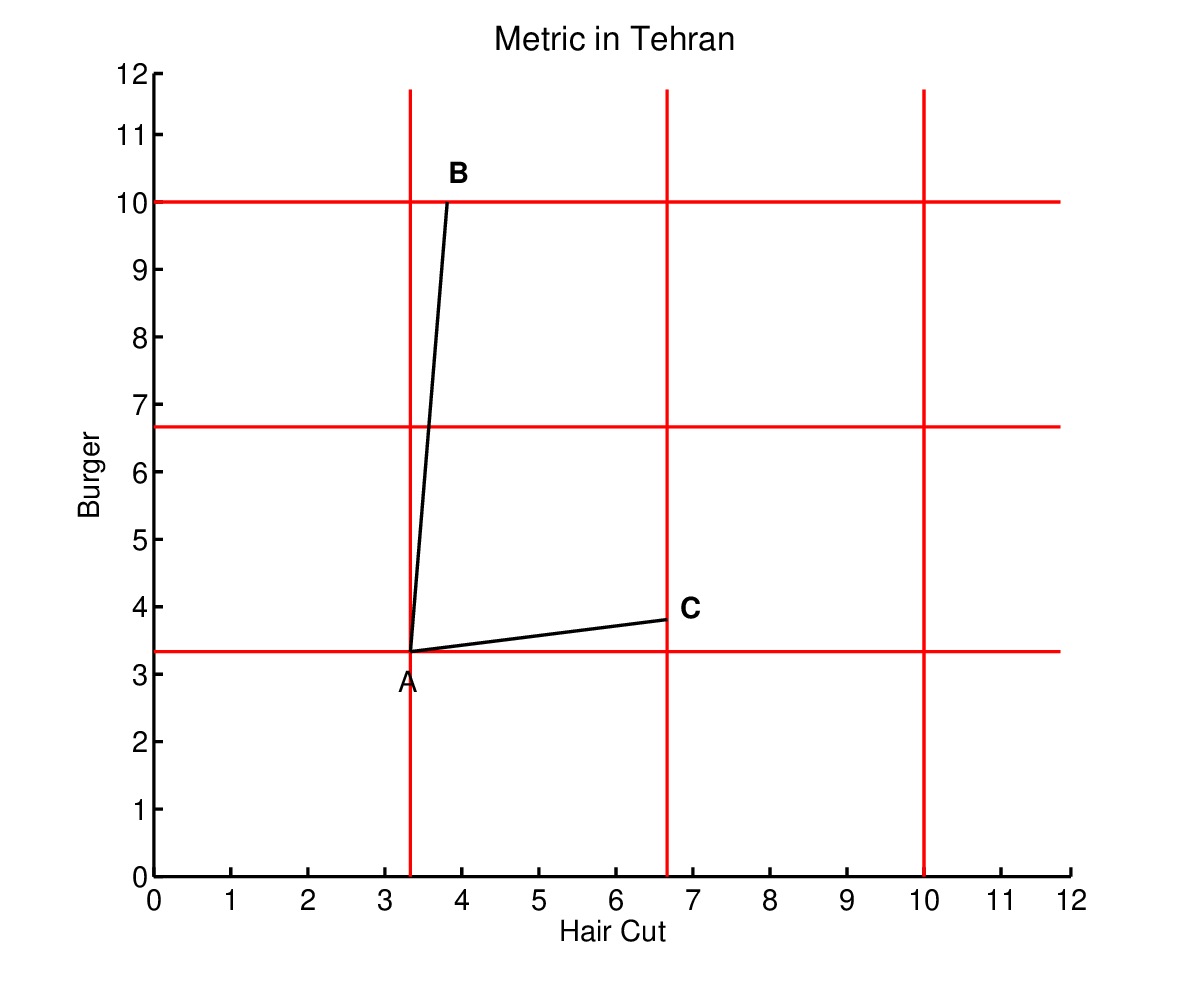}
                \label{figburgeriran}}
 \subfloat[][]{
    \includegraphics[width=0.65\columnwidth]{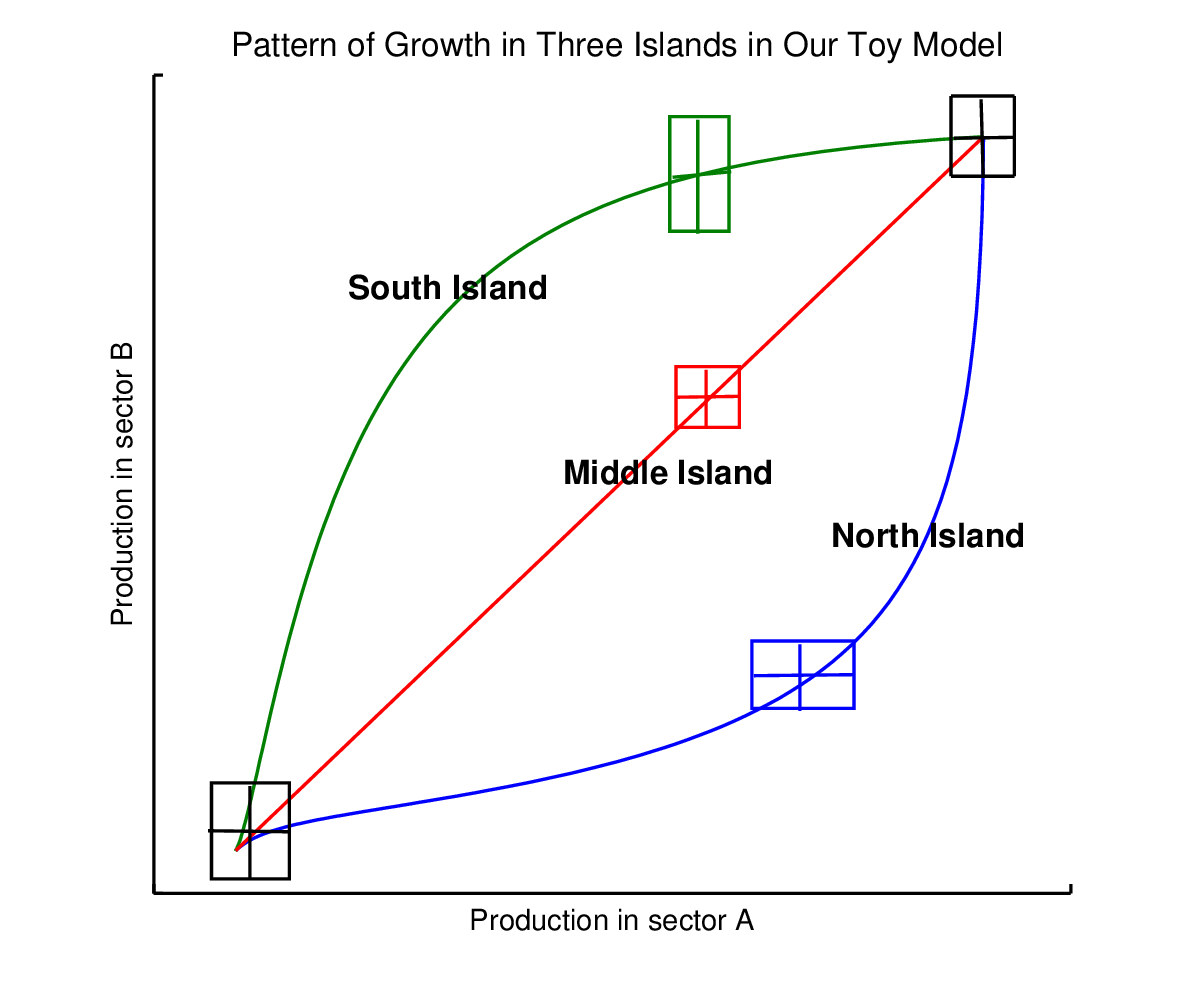}
                \label{figisland}}
\caption{ a) Price for a Mc'Donalds burger in NY is about 3\$ and suppose that average price for men haircut is 21\$. In each axis distance from origin shows the number of burgers and haircuts. Grids however show dollar values. Distance between parallel red lines in either directions is 10\$. In NY money value of distance of A to C is much greater than distances of A to B. b)In Iran services are much cheaper and distance of A to C is smaller than A to B. When defining money value of distances in the space of productions, the related data provider of each country has its own meter in each direction. So, while in Iran moving from A to B is recognized to have a bigger effect on the real GDP, in the US moving from A to C has a bigger effect on the real growth. c)In our problem for three islands, each island moves along its own path and national agencies have their own meters to measure distance between initial and final points. So what they measure as the real gap between initial and final sizes of GDP is different from each other.}
\label{figgrowthes}
\end{figure*}

In New York city price for a McDonald's "Triple Cheeseburger" is 3\$. Let's suppose that the average of a haircut for men is about 21\$. Now we can draw a framework as of Fig. \ref{figburgerus}. In this framework distance from each axis shows production level. Red grid however represents dollar values. Distance between parallel lines in the red grid represents 10\$. If you move along haircut direction your influence in GDP is bigger from moving in burger direction. One step along haircut direction passes more than two lines and has an influence more than 20\$. Along burger direction however you need to have more than 7 steps to have such effect in the growth of GDP. 

In Iran price for both burger and men haircut is around 3 \$. In this case, in spite of the US moving along haircut direction has not a bigger influence on GDP size. While based on the meter of the US Bureau of Economic Analysis (BEA) distance between A to C is much longer than A to B, in Iran based on the meter of the central bank, A to C is shorter than A to B. 

For our toy model, each country in each year has its own technology and productivity level. Relative prices are different in each country. Though each of three islands have the same initial and final conditions, since their paths are different in the middle ages their measure for the growth during the period of 99 years is quite different, see Fig. \ref{figisland}. Metric to aggregate production and measure the real growth is different in each country.

It should be noted that in this model we supposed that all Islands have the same utility preference and thereby whenever their technologies were the same their productions were the same as well. So, they ended up in the same point in the space of production. In real world however in each country people have their own utility preferences. As a result even if they start from the same initial conditions they would follow different paths in the space of productions and they barely may meet each other in this space.

If we aim to list the counterparts of elements of the general relativity to the current discussions we can count as follows.\\\\
{\bf -} In the general relativity metric, denoted by $g_{\mu\nu}$ helps to measure distances in the space time along a curve $c$ through the relation
\bea\label{distance}
d(x_i,x_f)=\int_c \sqrt{g_{\mu\nu}dx^{\mu}dx^{\nu}}.
\eea
In economics prices are our meters and the gap between economies can be addressed via
\bea\label{gapgeommery}
\Delta GDP=\int_c \Sigma_aP_adY_a.
\eea\\
{\bf-} In the general relativity metric is obtained through variation of the Einstein-Hilbert action
\bea
S=\frac{1}{16\pi}\int d^dxdt\sqrt{-g}(R+ matter\;term).
\eea
 In economics, prices are addressed through interactions of sectors and actually equations (\ref{newcapital}) and (\ref{newwage}). Though rate of return or real interest rate is addressed from the utility preference.\\\\
{\bf -} In the general relativity objects move along geodesics and the extremums of distances in Eq. (\ref{distance}). In economics, paths of countries in the space of production can be addressed through maximizing utility preferences in equations (\ref{lagrange}) and (\ref{timeutility}).\\\\ {
{\bf Mathematical precision}\\\\
Before closing this section I should notify that finding distance between points in economics and physics are conceptually similar. Though we borrowed the concept of metric from the general relativity to notify the impact of measurements by different observers, the equations we  reviewed however do not meet the true mathematical definition of metric. Equation (\ref{gapgeommery})
resembles a one form. Mathematically it resembles the work done on an object in a force field. The work done along each path however depends on the path since $P_a$ changes when level of productions changes. Though mathematically the problem is closer to the problems with work and energy, the concepts of measurements and meters for aggregation, dependency of measurement to the path, and dependency of paths to the optimizations and local extremums 
are conceptually close to the observations in the general relativity. 


\section{The Gap Between Economies of China and the United States}

Before paying attention to the gap between economies of China and the US let's recall paradoxical results concerning Pen World Table and International Comparison Program. When you use purchasing power parity exchange rate to measure GDP of different countries, it means that for all of them you have used the same metric to measure their GDP, see Fig. \ref{figpppa}. The price base in this method is standard worldwide. In developed countries such as the US since productivity is very high, wages are relatively high and services are relatively more expensive. This is while in poor countries relative prices for services are low comparing to the US. So, there will be problem when we aim to compare growth rate of different countries. Systems of national account tries to provide a standard basis worldwide. Currently however many countries such as the US or China do not follow such system. These countries use their own price base. From the general relativity point of view, national agencies resemble falling observers in a curved space-time who have their own meters. It is while international agencies have standard meters and resemble an outside observer. Though system of national accounts and PPP exchange rates aim to provide a better international basis for aggregation. Still, space time inconsistency is observed. For more studies see \cite{Crawford}-\cite{oulton2014}. 

\begin{figure*}%
\centering
  \subfloat[][]{
    \includegraphics[width=0.65\columnwidth]{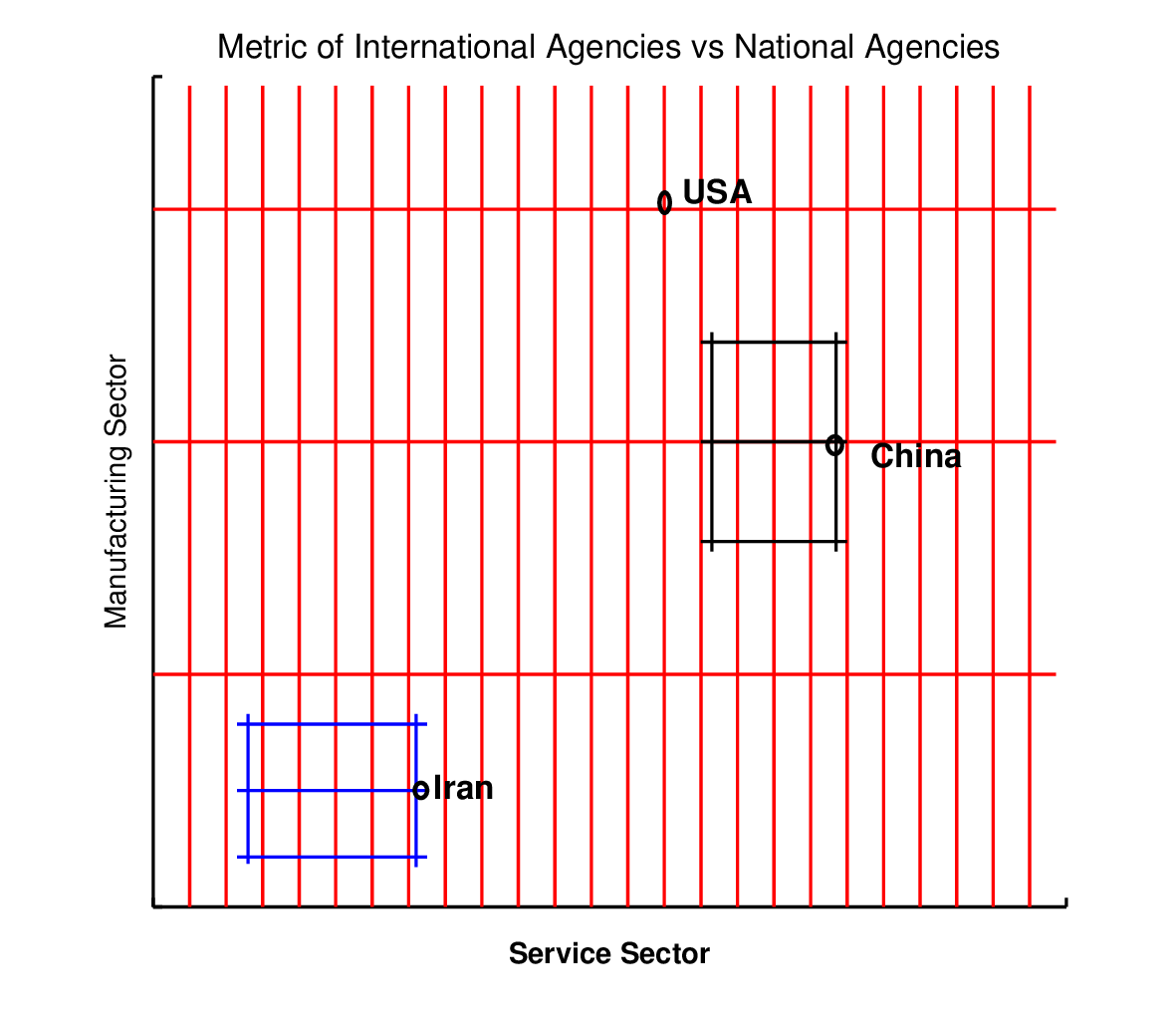}
                \label{figpppa}}
 \subfloat[][]{
    \includegraphics[width=0.65\columnwidth]{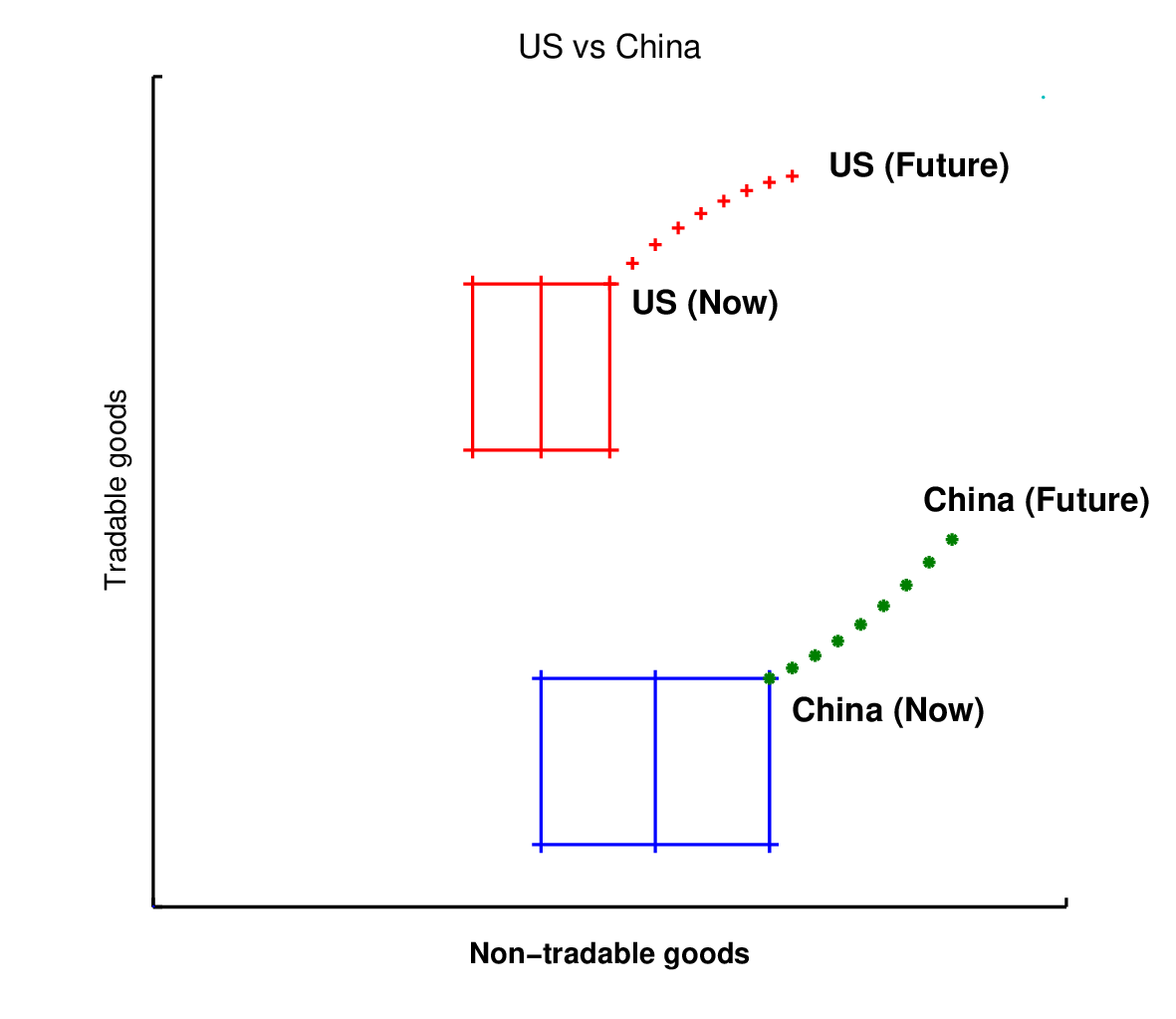}
                \label{figchinaa}}
 \subfloat[][]{
    \includegraphics[width=0.65\columnwidth]{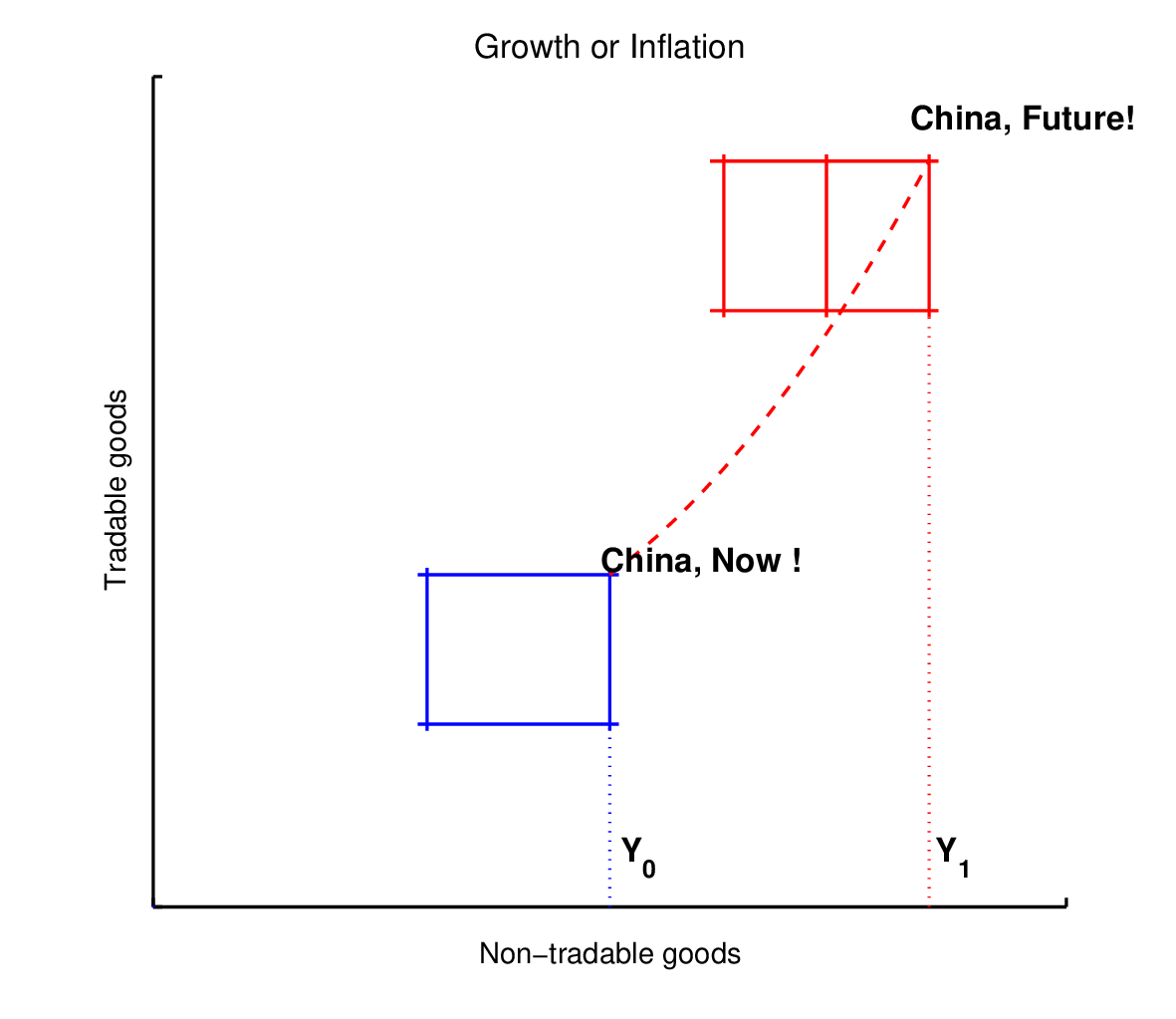}
                \label{figchinab}}
\caption{ a) While each country has its own prices or its own grid to measure different movements in real GDP, a PPP based measurement (red grid) aims to have a standard metric to measure GDP of different countries in the space of productions. b) The US and China have their own meter to measure GDP. While their meter to measure tradable goods are close, their meter to measure non-tradable goods are different. Each country pass its own path in the space of production and their meter specially along the non-tradable goods changes. c) As GDP in China grows, its meter along non-tradable goods changes and become closer to the ones for developed countries. While via new girding (the blue one), the nominal value of $Y_0$ has grown, from China NBS perspective only $(Y_1-Y_0)P_0$ is the real growth. From international perspective however meter of China to measure non-tradable goods has become closer to the international meter and the growth in the volume in non-tradable goods looks real. In other words from international perspective the volume of GDP of China has grown as $(Y_1-Y_0)P_1-Y_0P_0$.}
\label{figppp}
\end{figure*}

Back to our problem with China it is now clear that if we even know the current size of the GDP of both China and the US and besides we know their future growth rates we still are not able to examine the time that economy of China catches the economy of the US. To extrapolate the time that China needs to pass economy of the US besides knowing the future growth rates of both countries we need to know the path that each country passes through. We need to find this path through production functions and utility preferences and the level of capital in China and the US and then find local metric along this path and thereby find out the desired time. Since utility preferences in China is different from the US for sure the final shape of economy of each country will be different and thereby China should not meet the US in the space of production, see Fig. \ref{figchinaa}. If we know the path and local metric however we can extrapolate the time that nominal sizes of their GDP are the same.    \\\\
{\bf How dose it work?}\\

One may argue that we have nominal GDP of China and of the US. We have their reported real and nominal growths. How then Eq. (\ref{equation}) does not hold?  For sure the equation holds for nominal rates. The problem arises when we aim to extract real growth from the nominal growth. Each country reports its growth rate and then some day their nominal GDP sizes should be the same. The geometrical figure however is different. Tradable goods have roughly the same prices at border of both countries, see Fig. \ref{figchinaa}.  Non-tradable goods do not have the same price however. So, while girding of both countries along vertical direction roughly has the same length, their girding along horizontal axis is different. 
If we think of Baumol's cost disease and real observations, we find that prices for services are relatively cheaper in developing countries.

Manufacturing sectors have good overlap with tradable goods. It is while service sectors have better overlap with non-tradable goods. As time goes by, according to the cost disease and observations, relative prices of non-tradable goods increase. As a result girding of countries along horizontal axis will be more dense. Let's look at the case of China in Fig. \ref{figchinab}. What will happen then?

Before the period, production of both countries in tradable goods have had close prices. In non-tradable sectors however since prices in China have been smaller than the US, then nominal size of these sectors in China has been smaller respect to the US one. 

When economy of China grows along tradable axis we have no problem. Tradable goods are gridded worldwide roughly with the same price. The problem is with non-tradable goods or in some senses services. For these sectors, after a while prices of services in China will be closer to the prices of the US if we believe in the cost disease. What would happen then? For these sectors according to the National Bureau of Statistics of China (NBS) we have a growth in GDP size as:
\bea
\Delta GDP=Y_1P_f-Y_0P_i.
\eea
This is nominal share of growth for non-tradable sectors. The NBS however report part of this growth as inflation. According to the NBS the real growth is
\bea
\Delta GDP=(Y_1-Y_0)P_i.
\eea
What does it mean? China considers only the part of the growth that is real in this direction for the real growth rate. What about international comparisons. In international comparisons the nominal share of non-tradable goods has grown and from international perspective economy of China has grown. 
At the end of the period when GDP of China grows and China faces inflation along non-tradable goods, its price gets closer to the US one, nominal volume of the non-tradable goods grows. Though China considers parts of these growth as inflation, but since prices for these sectors have really become closer to the US one, and nominal GDP of China has become closer to the US one, from international comparisons it is a real growth. 

\begin{table*}
\begin{ruledtabular}
\begin{center}\begin{tabular}{ l c c  c c c  c   r }
Year \;& Country \;\;& \# of Good A \;& Price \;\; &\# of Service B \;&Price\;&Nominal GDP\;\\ \hline
2015& USA & 2000 & 1\$ &200 &10\$&4000\$\\ \cline{2-7}
 2015& China & 2000 &1\$
&300&5\$&3500\$\\ \hline\hline
2016  & USA&2000& 1\$ &200&10\$& 4000\$\\ \cline{2-7}
2016&  China  & 2120 &1\$&303&5.15\$
&3680\$\\ 
\end{tabular}\end{center}
\end{ruledtabular}
\caption[tab1]{Hypothetical production and prices in the US and China.}\label{tabb1}
\end{table*}

In international comparisons tradable goods have worldwide prices. Now that volume of services in China grows, then nominal GDP of China will be closer to the US one. It does not matter if China percepts it as inflation or real growth rate. To clarify it I have provided a numerical example in table \ref{tabb1}. 

I have supposed that the US and China produce only one good and provide only one service. Prices and productions are as the level in the table. In this hypothetical model from 2015 to 2016 nothing changes in the US. So, both inflation and growth experience a zero value rates. In China tradable good A has the same value as of the US. Service B however experiences a relative growth in prices in China since economy in this country is growing faster than the US. For this case the China NBS reports a real growth as of
3.9\%. According to the NBS part of the growth of the volume in service sector is inflation rather than being a real growth. So, China report an annual inflation equal to 1.3\%. So, if we consider reports of the NBS we should conclude that economy of China has become closer to the economy of the US by 3.9\%. If we look from an international perspective things are different. Growth of prices in service sector in China is not a fluctuation. It is permanent and is a result of the growth and the cost disease. Volume of economy of China has grown by 5.2\% from an international perspective. Meter of the NBS of China has become closer to the meter of the US BEA. From an international perspective volume of GDP of China has become closer to the US one by 5.2\% despite reports of the national agencies for the real growths.

Two conclusions can be deduced from our discussions: The first conclusion is that extrapolating the real time that nominal GDP of China will be equal to the nominal GDP of the US is a hard task even if we know their real growths. To find the proper answer we need to know both the relative growth of prices as a result of the cost disease and as the utility preferences. The problem holds with both nominal and PPP values of GDP sizes. 

We have two forms of variation in prices. One form of variation is the regular fluctuations of prices in the market as a result of short time interaction in supply and demand curves. The other form of variation is the growth in prices in service side as a result of the cost disease which is permanent as long as we can not find a technology to grow productivity. Though from national perspective growth in prices in service sector is inflation, from international perspective it is a permanent growth of volume of service sector and a permanent growth of the volume of GDP. 

Prices for non-tradable goods are lower in developing countries. If developing countries experience fast growth and relative prices for their services grow faster than manufacturing side, then this growth only make nominal GDP for them closer to the PPP values. Currently PPP value of GDP of China is bigger than the US. When GDP of China expands we expect prices in the service side in China become closer to the prices in the US. As a result, nominal size of GDP of China become closer to its PPP value. So, the second conclusion is that if prices for services in China become closer to the US ones then its economy become closer to the economy of the US sooner than what an equation such as Eq. (\ref{equation}) suggests.
 
\section*{Acknoledgments}

I would like to thank GR Jafari for comments on an earlier draft and H Arfaei for discussion. 


\end{document}